# Aging and the Narrowing of Scientific Innovation

## Aging researchers and the removal of retirement policies yield decreased disruptive innovation in science

**Authors:** Haochuan Cui[1,2,3,5],Yiling Lin[1], Lingfei Wu[1]*, James A. Evans[4,5,6]*

**Affiliations:**

[1] School of Computing and Information, University of Pittsburgh, Pittsburgh, PA 15260

[2] School of Journalism and Communication, Nanjing Normal University, Nanjing 210097, China

[3] School of Systems Science, Beijing Normal University, Beijing 100875, China

[4] Department of Sociology, University of Chicago, 1126 E 59th St, Chicago, IL 60637.

[5] Knowledge Lab, University of Chicago, 5735 South Ellis Avenue, Chicago, IL 60637.

[6] Santa Fe Institute, 1399 Hyde Park Road, Santa Fe, NM, 87501.

*Corresponding authors. E-mail: liw105@pitt.edu (W.L.) and jevans@uchicago.edu (J.E.)

Scientific careers today are marked by growing polarization. A small number of scientists now remain active and influential for longer than ever (*1*), while many others pass through research as temporary workers (*2*). Lengthened training periods (*3*, *4*), the elimination of mandatory retirement (*5*), and funding systems that reward experience have concentrated resources among senior scientists (*6*). As science becomes increasingly dependent on its aging core, a central question arises: *How does academic age influence creativity?*

The answer has long divided scholars. Early studies suggested a decline in creativity with age (*7*, *8*), while others found little or no relationship (*9–12*). Classic life-cycle analyses of scientific productivity and creativity likewise reveal diverse age patterns (*10*, *13*, *14*). One view holds that longer careers and deeper expertise help scientists keep pace with expanding knowledge. For example, the “burden of knowledge” theory argues that familiarity with old ideas enables new ones to emerge (*15*). Yet empirical evidence often shows the opposite: innovation tends to come from younger scientists who are less tied to convention—echoing the gloss of Max Planck’s insight that “science advances one funeral at a time” (*16*, *17*).

A likely resolution to this paradox lies in how aging reshapes the balance between two kinds of creativity. Familiarity with established ideas may expand a scientist’s capacity for recombination—creating novel links between known concepts—but also strengthens attachment to these ideas, making them harder to abandon (*7*, *18*). In other words, experience aids recombinatorial novelty (*19*) but limits disruptive innovation (*20*), which challenges prior work and drives the creative destruction of ideas (*21*).

History offers vivid examples. During periods of scientific revolution (*22*), the weight of accumulated knowledge can slow older scientists, while younger minds who remain freer from convention, race toward the frontiers of discovery. Even the greatest are not immune. Albert Einstein reshaped our understanding of space, time, mass, and energy at twenty-six, yet spent much of his later career struggling—without success—to unify his earlier theories (*23*) and resist the rise of quantum mechanics (*24*). Ironically, his critiques helped sharpen and solidify the very framework he opposed, propelling the quantum age forward (*25*). The same rhythm appears beyond science: in art and literature, young "revolutionaries" spark breakthroughs while older "masters" refine traditions (*26*), and in law, mandatory retirement for judges improves court performance (*27*).

We test this pattern of intellectual aging across modern science by distinguishing between two empirically measured forms of creativity. *Novelty* links previously unconnected ideas, applying established knowledge in new contexts (*19*, *28*, *29*). *Disruption* breaks existing linkages, displacing established knowledge with new ideas (*30–32*). Although both drive progress, they reshape scientific knowledge in opposite ways: novelty extends established paradigms by revealing hidden complementarities between ideas, whereas disruption overturns them by exposing hidden substitutions (*33*).

These empirical distinctions build on recent research, but their intellectual roots span long traditions in the study of science and technology. Theories of recombinatorial novelty trace back to Schumpeter (*34*), who defined innovation as the creation of "new combinations" of existing resources, a view later expanded by Mokyr (*35*), Weitzman (*19*), and Arthur (*36*). This view was empirically examined by Fleming et al. (*28*, *37*) in technology and by Uzzi et al (*29*) and others (*38*, *39*) in science. Theories of disruptive innovation likewise draw on both traditions: in studies of science, where Popper (*40*) emphasized falsifying and displacing existing ideas as a driver of progress, a process Kuhn (*41*) termed "paradigm shift" and Merton (*7*) called "obliteration by incorporation", further elaborated by subsequent scholars (*42–48*). In studies of technology, the concept originates in Schumpeter's (*49*) notion of "creative destruction", expanded by Aghion and Howitt (*21*) and Christensen (*20*). Building on these traditions, Funk and Owen-Smith (*30*) operationalized disruption in technology, and Wu et al. (*31*) introduced and validated this framework in science.

Analyzing more than 3.6 million scientists who published between 1960 and 2020, we find that novelty increases while disruption declines with academic age. We trace this divergence to a "*Nostalgia Effect*"—older scientists' tendency to cite earlier work. This preference for the past stimulates new combinations of familiar ideas, enhancing novel combinations, but it also anchors attention to established paradigms, reducing disruption and fostering criticism of new ideas proposed by younger colleagues. As scientists age, they also form more hierarchical teams, amplifying their reach yet reinforcing reliance on conventional knowledge. Policies such as the 1994 removal of mandatory university retirement in the United States, though intended to reduce age bias, may have inadvertently deepened this intellectual conservatism.

Nevertheless, our findings offer hope. Mobility and collaboration can slow intellectual aging. When scientists move between institutions or when younger researchers step into leading roles, such as serving as corresponding authors, the age of references—and of ideas—falls, refreshing the collective memory of science.

Our study contributes to three areas. First, it clarifies why prior research on age and creativity has been mixed (*7–12*): aging enhances combinatorial innovation but limits disruptive breakthroughs. Second, it links these cognitive dynamics to team structure, explaining why larger, hierarchical teams tend to innovate less (*31*, *50*, *51*), thereby revealing the social foundations of intellectual aging. Third, it identifies the Nostalgia Effect as a measurable cognitive process that shapes how scientists remember, connect, and—at times—struggle to forget ideas over their careers (*52–54*).

To examine these patterns empirically, we begin by identifying name-disambiguated scientists using the 2021 Microsoft Academic Graph (MAG), now part of the OpenAlex dataset. Our sample includes 7.7 million scientists who have published between 1960 and 2020, each with five or more publications. We verified name disambiguation by manually reviewing 1,118 papers from 30 randomly selected scientists, achieving 96.8% precision across two independent coders (see Supplementary Materials for additional validation tests).

To examine the dynamics of cognitive focus, we defined *academic age* as the number of years since a scientist's first publication and traced how their primary knowledge sources, captured by the *average reference age*, change over time. Average values can be distorted by outliers, such as when extremely old references inflate the mean (*7*) or prolific scientists lower it by citing more recent work (*12*). We developed a distribution-based measure that captures the full temporal profile of a scientist's *most-cited references*, reducing bias and preserving variation.

We assessed *novelty* by how papers draw upon previously unconnected domains, diverging from conventional citation patterns (*29*). We replicated these results using alternative novelty metrics that quantify the "surprise" of content and context combinations (*39*). We assessed *disruption* by how subsequent papers cite a focal paper but not its references, indicating when new ideas displace older ones (*55*). We also quantified *topic exploration* by comparing research topics across consecutive years of each scientist's publications, using field-of-study keywords from the MAG dataset (*56*) and repeating the analysis across multiple levels of keyword resolution to ensure robustness.

These analyses allow us to examine the life cycle of scientific aging, which we emphasize is not merely biological but deeply social. As scientists progress from trainees to principal investigators, their roles and constraints shift markedly. Increasing leadership, administrative, and reviewing responsibilities can reshape priorities and limit the time available to stay current with the research frontier, thereby slowing reference renewal. In this sense, individual aging also carries collective consequences. Within teams, senior scientists' preferences can shape the citation practices of their collaborators, especially those they lead. In peer review, older reviewers may direct authors toward citing familiar or preferred work, further reinforcing established literature. Together, these social mechanisms reveal how aging scientists shape the collective evolution of scientific memory.

To investigate team dynamics, we compared hierarchical teams—where older scientists lead and younger members follow (*57*, *58*)—with flat teams, where leadership is shared (*59*, *60*). We also examined how mobility affects reference aging by tracing when teams split or reunite across institutions. Using quasi-experimental comparisons of papers from the same teams before and after changes in corresponding authors or collaboration distance, we identified a causal link between team structure, collaboration distance, and the age of cited references.

To investigate peer-review dynamics (*7*), we analyzed changes in reference age between preprints and their published versions to examine how reviewer age, estimated from field-level averages, influences these changes. To understand the origins of individual reviewing effects, we further analyzed scientists' attitudes toward new ideas through *critical* *citations*. Building on prior work (*61*, *62*), we constructed a new, human-validated dataset of 20,000 critical citation examples. Manual validation of 200 randomly selected samples achieved 88% inter-coder agreement between two human coders. A large language model, validated with 86.3% precision, was then applied to 126 million citation contexts across 7.1 million papers in MAG to examine how the likelihood of citing or being cited critically varies with academic age. To rule out the possibility that the observed pattern reflects machine-learning model bias, we alternatively classified critical citations using a curated list of critical terms, yielding the same pattern of results (see SM S3.7.6, Fig. S14, Table S4).

Finally, to assess the broader impact of these mechanisms on the scientific system, we conducted a difference-in-differences (DID) analysis of the 1994 U.S. policy that eliminated mandatory university retirement. To our knowledge, this represents the first quasi-experimental study to quantify the social dimension of intellectual aging.

Science advances by building on recent discoveries—a pattern known as the immediacy effect (*63*), with most citations in an average paper drawn from the past decade. Our analysis confirms this trend, showing an average reference age of 9.8 years, ranging from 7.1 in computer science to 12.7 in mathematics (Fig. S1).

However, not all scientists adopt new literature at the same pace. As careers unfold, a *Nostalgia Effect* emerges as scientists increasingly cite older work. On average, the age of references increases by about one month (0.09 year) per career year, implying that by a 40-year career age, a scientist's reference list becomes roughly 3.6 years older, 37% of the global mean reference age (Table S1). This magnitude of change varies across fields: in medicine, the reference age rises by 17% (from 8.7 to 10.2 years) over 40 years, whereas in mathematics it increases by 50% (from 12.2 to 18.2 years; Fig. 1A). This pattern remains robust when using the Price index (*63*) the share of references less than five years old, which declines steadily with career age (Fig. 1A, inset). The pattern is consistent across career lengths and historical cohorts, with no evidence that either—or their interaction—modulates the effect (Fig. 1B, S2). We also rule out alternative explanations. The pattern is not driven by self-citations (Fig. S3A) or by a shift from empirical to theoretical work over a career (Fig. S3C). Working at top universities appears to slow aging (Fig. S3B), whereas early career success accelerates it (Fig. S3D). Overall, the aging of references is a universal feature of scientific careers.

The temporal distribution of scientists' most-cited references confirms this pattern and rules out artifacts, such as a few very old citations that inflate the average (*7*). More than a robustness check, the most-cited reference defines the foundational work anchoring a scientist's knowledge base: across the career, 96% of references appear only once, yet the most-cited one recurs in nearly half (48%) of their papers. Its early-career peak reveals an imprinting effect: the paper that most strongly shapes a scientist's career is typically published one year before their own first publication, regardless of career length or cohort (Fig. 1C, S4). Because breakthroughs occur irregularly across time (*22*), the fact that a scientist's most-cited reference peaks at career birth indicates imprinting rather than paper quality. Together, these findings highlight a universal tension between science's forward momentum and aging scientists' growing attachment to the intellectual past.

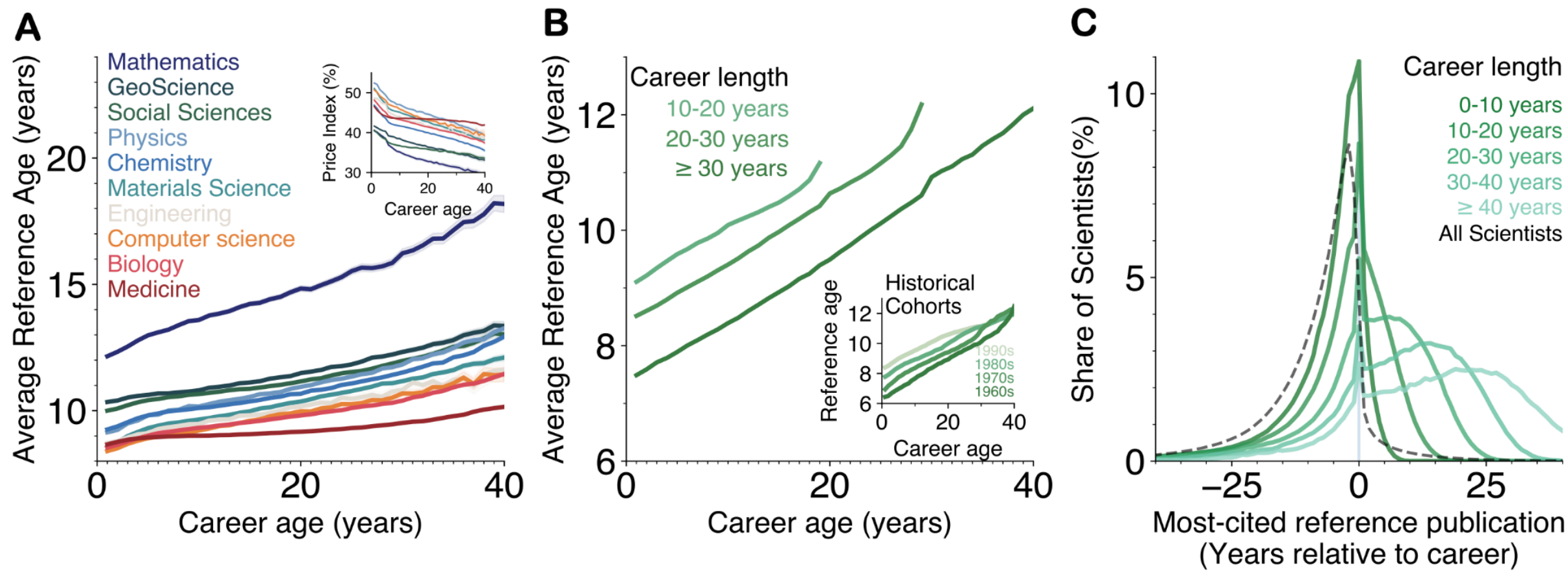

**Fig. 1: The aging of scientific memory. A**, For 7,740,433 scientists with five or more papers, the average age of references rises monotonically with career age—from 8.7 to 10.2 years in medicine and from 12.2 to 18.2 years in mathematics over four decades. The inset shows the Price Index (*63*), the share of references less than 5 years old, declining steadily with career age. **B**, This pattern holds across scientists with different career lengths and remains stable across historical cohorts (inset). **C**, Each scientist's most-cited reference—their intellectual anchor—typically appears near the start of their career but shifts progressively later as careers lengthen.

As scientists age, their creative orientation shifts. The probability of producing novel work increases steadily with experience (Fig. 2A). Alternative novelty metrics that capture the "surprise" of content and context combinations (*39*) show the same upward trend, indicating greater combinatorial creativity (Fig. S5). Yet this novelty largely stems from linking and recombining familiar ideas within specialized domains rather than venturing into new ones. Exploring new topics, measured by the cosine distance between publication keywords in consecutive years, declines sharply with career age (Fig. 2A, inset), a pattern robust across levels of keyword granularity (Fig. S6). In network terms (*64*, *65*), aging scientists add edges—new connections among existing concepts—rather than nodes—entirely new concepts—within the conceptual space of science. This shift from expanding the network's boundaries to connecting its interior captures how experience favors recombinant creativity over exploratory or disruptive innovation.

In contrast, the likelihood of producing highly disruptive papers decreases over the career (Fig. 2B), indicating that older scientists become less likely to overturn established ideas. This life-cycle pattern holds across research fields, historical cohorts, and career lengths (Fig. S7), as well as across alternative thresholds and citation windows for disruption (Fig. S8-9), suggesting that it reflects cognitive and social shifts within scientific careers rather than differences in citation practices (*66*, *67*).

Cross-national comparisons reinforce this pattern. Countries with younger scientific workforces, such as China and India, produce a larger share of disruptive papers, whereas those with older workforces, such as the United States, exhibit lower disruption rates (Fig. 2C). Together with the Nostalgia Effect, these results suggest that familiarity with established ideas narrows the scope of creative search: aging scientists grow more adept at recombining what they know but less inclined (or able) to abandon old concepts and replace them with new ones.

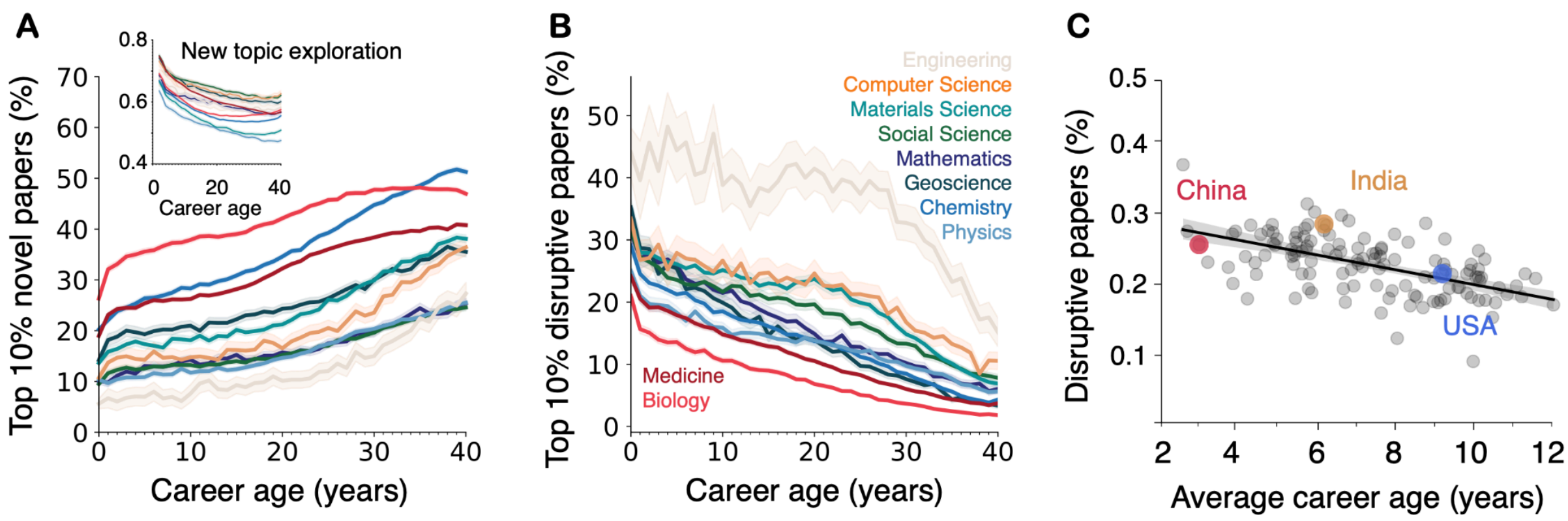

**Fig. 2: Innovation changes with career age. A,** Among 219,597 scientists with careers spanning 40 years or more, the probability of producing a top 10% *novel* paper increases with career age. Yet new topic exploration, measured by the cosine distance between publication keywords in consecutive years, declines with career age (inset). **B,** In contrast, the probability of producing a top 10% *disruptive* paper decreases sharply with career age across all fields, indicating that older scientists are less likely to overturn established ideas. **C,** Across 126 countries, the share of disruptive papers in 2020 declines with the average career age of active scientists: nations with younger scientific workforces (e.g., China, India) show higher disruption, whereas those with older ones (e.g., the United States) exhibit lower levels.

As academic age increases, scientists take on larger roles in science, shifting from individual contributors to mentors of junior scholars, leaders of teams and institutions, and reviewers of papers and grants (*7*). Through these social roles, they shape science's collective memory, evaluative standards, and innovation potential beyond the influence of their own publications. Consequently, the Nostalgia Effect extends from individual cognition to a system-level pattern that determines which ideas endure, gain credit, and inspire new work.

We find that the average age of a team's references increases more rapidly with team age than individual age, diverging from a neutral baseline that assumes equal influence from all members. This widening gap suggests that older members—especially team leaders—pull citations toward older literature beyond what simple averaging would predict (Fig. 3A).

Team age dynamics reveal how this influence emerges. As the scientific workforce becomes increasingly polarized by career age (Fig. S10), teams commonly mix junior and senior scientists but are increasingly led by older ones (Fig. 3B, inset). As focal scientists age, the average age of their coauthors also increases, though at a much slower rate—typically by ≤15 years over a 40-year career (Fig. 3B, S11). Older leaders thus anchor mixed-age teams and steer collective attention toward earlier literature.

A quasi-experimental test confirms that collaboration patterns shape team cognition (*68*) and collective aging. Among the same teams publishing at different times, later papers led by younger corresponding authors cite significantly newer references than those led by older ones. Teams that merge from remote to onsite collaborations also cite newer work, whereas those that split into remote groups cite older work (Fig. 3C). Together, these within-team contrasts indicate that the influence of aging scientists on collective memory depends on team design and that breaking age and power hierarchies fosters team innovation (*50*, *51*, *69*).

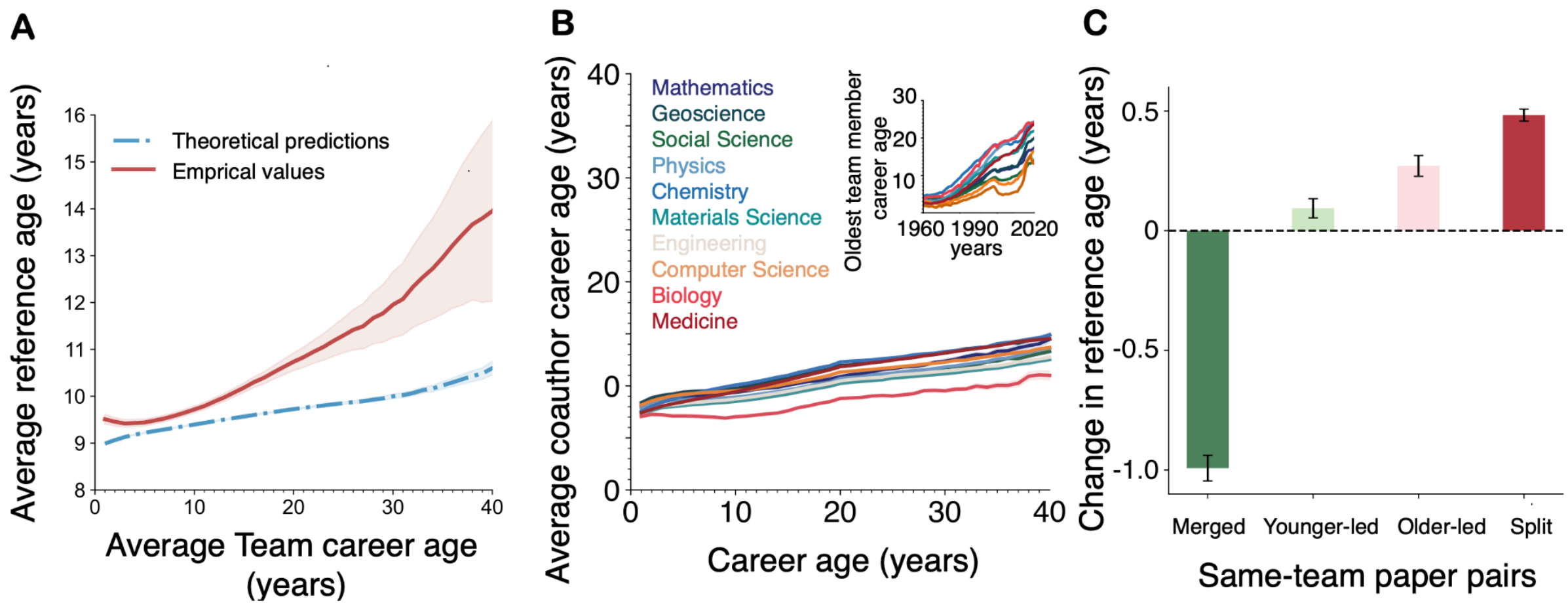

**Fig. 3: The influence of aging scientists on teams**. **A,** Average reference age rises with a team's average career age and exceeds a neutral baseline that assumes equal influence among teammates. **B,** As focal scientists age, their coauthors get older—but much more slowly: over a 40-year career, the average coauthor age rises by only about 15 years. Meanwhile, the oldest team member's career age has increased in recent decades (inset). **C,** Same-team paper pairs link team changes to citation freshness: among 383,282 pairs with mobility shifts, teams merging from remote to onsite (n = 92,979) cited newer work, whereas those splitting to remote (n = 290,303) cited older work (95% CIs shown). Among 176,366 pairs with changes in corresponding-author age, later papers generally cited older work, but those listing younger corresponding authors (n = 96,589) cited significantly newer references than those listing older ones ($n = 79{,}777$; $t = -5.86$, $p < 0.001$).

Aging scientists also shape science through their evaluative roles. Across ten fields, including mathematics, computer science, and physics, the likelihood that a journal version of a paper cites older references than its preprint increases with the average career age of reviewers (logistic regression coefficient = 0.19, $p < 0.001$), with reviewer age estimated from field averages (Fig. 3A).

As scientists age, their attitudes toward new work also shift. The share of papers that criticize their cited references rises with career age, stabilizing around ~20% across fields (Fig. 3B, Fig. S14, Table S4). In contrast, the share of papers that receive critical citations declines with author age (Fig. 3B, inset), suggesting that older scientists are more likely to critique younger colleagues than to be challenged by them. Further analysis confirms that these critical citations are associated with the Nostalgia Effect—authors are more likely to dispute newer discoveries when recalling and citing their most-cited references over careers (*70*). To ensure critical attitude is not driven by older scientists' reactions to low-quality work, we show that as scientists age, they are more likely to criticize the 581 papers that eventually received Nobel Prize recognition as well as "Sleeping Beauty" papers that eventually receive substantial recognition despite being overlooked early on (Fig. S12-14).

Policy evidence aligns with the Nostalgia mechanisms. Following the 1994 U.S. Supreme Court decision that ended mandatory university retirement, reference ages rose significantly in the United States relative to the United Kingdom, used as a same-year baseline (Fig. 3C). This natural experiment indicates that reduced turnover among senior scholars increases reliance on older literature. Placebo and robustness tests confirm that this effect is not driven by coincidental temporal trends (Fig. S15). Regression analyses controlling for time period and field further rule out the possibility that the observed increase in reference age results from the growing availability of digital archives (Table S2) (*71*, *72*).

Consistent with these aging patterns in memory, field-level aging is associated with collective consequences for innovation. As a field's scientists grow older on average, mean reference age within that field increases even after accounting for individual career ages (Table S3). Moreover, as fields age, the emergence of new topics declines approximately linearly, indicating that workforce age structure shapes the pace of innovation (Fig. S16).

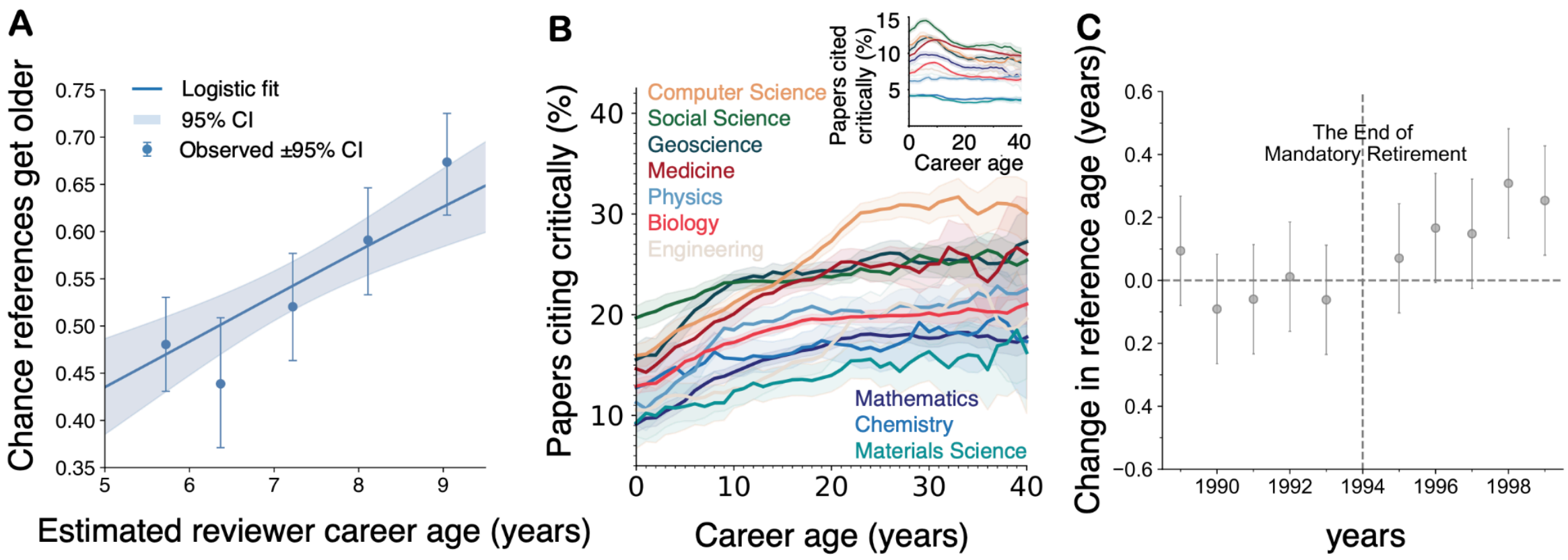

**Fig. 4: The influence of aging scientists on science. A,** For 1,447 papers across ten research fields, the likelihood that journal publications cite older references than their corresponding preprints increases with the average career age of reviewers (coefficient = 0.19, SE = 0.041, $z = 4.73$, $p < 0.001$). Reviewer age is estimated from field averages. Error bars represent 95% confidence intervals. **B**, Share of papers that criticize other work rises with career age across fields. In contrast, the share of papers receiving criticism increases early in the career but gradually declines over the long run (inset). The dataset includes over 9 million scientists who authored 7 million papers containing 126 million citation contexts. **C,** Estimated impact of the 1994 U.S. removal of mandatory academic retirement on reference age, based on 2,229,239 papers authored by U.S.-based scientists and 550,932 papers authored by U.K.-based scientists between 1989 and 1999. Each dot represents a coefficient from fixed-effects regressions of reference age on year indicators. Reference ages rose significantly after 1994 relative to the U.K. baseline (dashed).

## Discussion

We provide quantitative evidence for a Nostalgia Effect in science: as academic age increases, scientists' attention shifts toward older literature. This pattern reflects cognitive and social processes beyond biology alone—scientists' reactions to new ideas depend less on idea content than on their own career stage (*7*). Familiarity with established work fosters attachment, making replacement harder (*33*). Even the greatest minds, such as Einstein, transitioned from disruptor to gatekeeper as the "burden of knowledge" mounted (*15*).

This life-cycle pattern accords with science fiction author Douglas Adams' observation about acceptance of new technologies—what exists at one's intellectual "birth" feels normal, what appears during early career feels revolutionary, and what emerges after maturity feels suspect (*73*). Similar attachment dynamics appear in language, music, and the arts (*8*, *26*, *52*, *53*, *74–76*), reflecting shared long-term memory mechanisms across creative domains.

Critically, role transitions magnify individual tendencies. As careers advance, scientists shift from producing research to leading teams and institutions, reviewing papers and grants, and mentoring students (*7*). These social roles shape science's collective memory and evaluative standards, channeling attention toward established ideas and away from conceptual replacements. What appears as a decline thus also reflects functional reorientation: from overturning paradigms to consolidating knowledge and governing the conditions under which younger scientists pursue disruptive innovation.

Our findings carry several policy implications, which we frame as reflective rather than prescriptive. First, the decline in disruption with academic age challenges the assumption that experience alone drives frontier-moving advances (*11*, *15*). Senior scientists excel at recombining existing ideas, but their reduced propensity for paradigm change calls for more balanced funding and promotion systems. These results do not suggest that older papers should not be cited. Many classic studies and "sleeping beauties" (*77*, *78*) continue to shape modern research. Rather, they reveal a long-term memory mechanism in scientific practice: researchers tend to recall and cite papers encountered early in their careers, reflecting confidence built through familiarity and repeated use. This mechanism sustains the continuity of ideas across generations, yet it can also constrain the emergence of innovations that displace established knowledge. Because scientific progress depends on both continuity and renewal, policies should promote diverse pathways that enable both, especially as the global research workforce continues to age.

Second, the 1994 U.S. removal of mandatory university retirement serves as an illustrative case of how well-intended policies can reshape knowledge use. After the reform, reference ages rose relative to the baseline, consistent with reduced turnover in fields with a higher concentration of senior scholars. Recent research reports a parallel pattern in the U.S. legal system, where mandatory retirement is linked to productivity and influence (*27*). Science policy should therefore assess how funding mechanisms, tenure structures, and retirement norms together shape the age distribution of fields and, in turn, patterns of innovation.

Third, age structure matters for international competition. Countries with younger scientific workforces (e.g., India, China) tend to produce more disruptive work, whereas those with older workforces (e.g., Japan, the United States) tend to integrate and recombine knowledge (*1*). These cross-national patterns are descriptive rather than causal and serve to situate our findings within broader demographic trends. In the U.S., immigrant scientists—younger on average and concentrated in fast-growing STEM areas—mitigate aging trends and bolster capacity (*79*, *80*). Strategic investments in workforce development should leverage these complementary strengths and rebalance funding toward higher-risk projects led by early-career researchers (*81*).

Finally, team-level actions are immediately actionable. Our results show that team dynamics transmit the Nostalgia Effect: older leadership shifts citations toward older literature, whereas younger corresponding authors and mobility refresh what teams read and cite. Institutions can counter this narrowing by expanding early-career PI and co-corresponding author roles, encouraging intergenerational, flat collaborations (*50*, *51*, *82*), increasing collaboration diversity (*83*), lowering barriers to mobility and cross-institutional work (*84*), and recognizing disruptive contributions alongside recombinational synthesis.

By acting on the social mechanisms that channel individual aging into what gets funded, published, and cited, these steps keep science open to disruption while harnessing the integrative strengths of experience.

**Supplementary Materials:**

[https://docs.google.com/document/d/1clSEkEEeMidxmtz3KAlLz69kxRzuoceLRKoSijf5sDg/edit?usp=sharing](https://docs.google.com/document/d/1clSEkEEeMidxmtz3KAlLz69kxRzuoceLRKoSijf5sDg/edit?usp=sharing)

**Data Availability**

The datasets used in this paper are available at https://hashc.github.io/AgingScience/. Source data are provided with this paper.

**Acknowledgments**

We acknowledge support from the National Science Foundation (grants 2239418 for L.W., and 1829366 and 1800956 for J.A.E.), the National Institutes of Health (R01GM164731 for L.W.), and the Air Force Office of Scientific Research (FA9550-19-1-0354 and FA9550-15-1-0162 for J.A.E.). We thank the organizers and participants of the National Bureau of Economic Research (NBER) conference, *Investments in Early Career Scientists*, and the International Conference on the Science of Science and Innovation (ICSSI) for helpful feedback on earlier versions of this work.

**Contributions**

L.W. and J.A.E. jointly conceived and designed the study, interpreted the results, and drafted and revised the manuscript. H.C. and Y.L. conducted the data analysis and implemented the modeling framework.

**References**

1. D. M. Blau, B. A. Weinberg, Why the US science and engineering workforce is aging rapidly. *Proc. Natl. Acad. Sci. U. S. A.* **114**, 3879–3884 (2017).

2. S. Milojević, F. Radicchi, J. P. Walsh, Changing demographics of scientific careers: The rise of the temporary workforce. *Proc. Natl. Acad. Sci. U. S. A.* **115**, 12616–12623 (2018).

3. P. Stephan, J. Ma, The increased frequency and duration of the postdoctorate career stage. *Am. Econ. Rev.* **95**, 71–75 (2005).

4. S. Kahn, D. K. Ginther, The impact of postdoctoral training on early careers in biomedicine. *Nat. Biotechnol.* **35**, 90–94 (2017).

5. O. Ashenfelter, D. Card, Did the elimination of mandatory retirement affect faculty retirement? *Am. Econ. Rev.* **92**, 957–980 (2002).

6. T. Bol, M. de Vaan, A. van de Rijt, The Matthew effect in science funding. *Proc. Natl. Acad. Sci. U. S. A.* **115**, 4887–4890 (2018).

7. R. K. Merton, *The Sociology of Science: Theoretical and Empirical Investigations* (University of Chicago Press, 1973).

8. D. W. Galenson, B. A. Weinberg, Age and the quality of work: The case of modern American painters. *J. Polit. Econ.* **108**, 761–777 (2000).

9. S. Cole, Age and Scientific Performance. *Am. J. Sociol.* **84**, 958–977 (1979).

10. D. K. Simonton, Quality, quantity, and age: the careers of ten distinguished psychologists. *Int. J. Aging Hum. Dev.* **21**, 241–254 (1985).

11. B. F. Jones, B. A. Weinberg, Age dynamics in scientific creativity. *Proc. Natl. Acad. Sci. U. S. A.* **108**, 18910–18914 (2011).

12. S. Milojević, How are academic age, productivity and collaboration related to citing

behavior of researchers? *PLoS One* **7**, e49176 (2012).

13. P. E. Stephan, S. G. Levin, *Striking the Mother Lode in Science: The Importance of Age, Place, and Time* (Oxford University Press, New York, 1992).

14. D. K. Simonton, *Creativity in Science: Chance, Logic, Genius, and Zeitgeist* (Cambridge University Press, 2004).

15. B. F. Jones, The Burden of Knowledge and the "Death of the Renaissance Man": Is Innovation Getting Harder? *Rev. Econ. Stud.* **76**, 283–317 (2009).

16. P. Azoulay, J. S. Graff Zivin, J. Wang, Superstar Extinction. *Q. J. Econ.* **125**, 549–589 (2010).

17. J. G. Zivin, P. Azoulay, C. Fons-Rosen, Does Science Advance One Funeral at a Time? *Am. Econ. Rev.* **109**, 2889–2920 (2019).

18. W. B. Arthur, Competing technologies, increasing returns, and lock-in by historical events. *Econ. J. (London)* **99**, 116 (1989).

19. M. L. Weitzman, Recombinant growth. *Q. J. Econ.* **113**, 331–360 (1998).

20. C. M. Christensen, *The Innovator's Dilemma: When New Technologies Cause Great Firms to Fail* (Harvard Business Review Press, 2013).

21. P. Aghion, P. Howitt, A model of growth through creative destruction. *Econometrica* **60**, 323 (1992).

22. T. S. Kuhn, The structure of scientific revolutions: University of Chicago press. *Original edition* (1962).

23. A. Einstein, On the Generalized Theory of Gravitation. [Preprint] (1950). https://doi.org/10.1038/scientificamerican0450-13.

24. A. Einstein, B. Podolsky, N. Rosen, Can Quantum-Mechanical Description of Physical Reality Be Considered Complete? *Phys. Rev.* **47**, 777–780 (1935).

25. A. Pais, Einstein and the quantum theory. *Rev. Mod. Phys.* **51** (1979).

26. D. W. Galenson, *Conceptual Revolutions in Twentieth-Century Art* (Cambridge University Press, Cambridge, England, 2009).

27. E. Ash, W. B. MacLeod, Mandatory retirement for judges improved the performance of US state supreme courts. *Am. Econ. J. Econ. Policy* **16**, 518–548 (2024).

28. L. Fleming, Recombinant Uncertainty in Technological Search. *Manage. Sci.* **47**, 117–132 (2001).

29. B. Uzzi, S. Mukherjee, M. Stringer, B. Jones, Atypical combinations and scientific impact. *Science* **342**, 468–472 (2013).

30. R. J. Funk, J. Owen-Smith, A Dynamic Network Measure of Technological Change.

*Manage. Sci.* **63**, 791–817 (2017).

31. L. Wu, D. Wang, J. A. Evans, Large teams develop and small teams disrupt science and technology. *Nature* **566**, 378–382 (2019).

32. Y. Lin, L. Li, L. Wu, The Disruption Index measures displacement between a paper and its most cited reference, *arXiv [cs.DL]* (2025). http://arxiv.org/abs/2504.04677.

33. L. Li, Y. Lin, L. Wu, *Can Recombination Growth Lead to Scientific Breakthroughs?* (2024).

34. J. A. Schumpeter, The Theory of Economic Development... by Joseph A. Schumpeter, Translated from the German by Redvers Opie. (1934).

35. J. Mokyr, *The Lever of Riches: Technological Creativity and Economic Progress* (Oxford University Press, 1992).

36. W. B. Arthur, The nature of technology: What it is and how it evolves. (2009).

37. L. Fleming, O. Sorenson, Technology as a complex adaptive system: evidence from patent data. *Res. Policy* **30**, 1019–1039 (2001).

38. J. Wang, R. Veugelers, P. Stephan, Bias against novelty in science: A cautionary tale for users of bibliometric indicators. *Res. Policy* **46**, 1416–1436 (2017).

39. F. Shi, J. Evans, Surprising combinations of research contents and contexts are related to impact and emerge with scientific outsiders from distant disciplines. *Nat. Commun.* **14**, 1641 (2023).

40. K. Popper, *The Logic of Scientific Discovery* (Routledge, 2005).

41. T. S. Kuhn, *The Structure of Scientific Revolutions: 50th Anniversary Edition* (University of Chicago Press, 1962).

42. I. Hacking, *Representing and Intervening: Introductory Topics in the Philosophy of Natural Science* (Cambridge University Press, Cambridge, England, 1983).

43. R. N. Giere, Explaining Science, *University of Chicago Press* (1990). https://press.uchicago.edu/ucp/books/book/chicago/E/bo3622319.html.

44. P. Kitcher, The advancement of science: Science without legend, objectivity without illusions. (1993).

45. L. Laudan, Progress and its problems: Towards a theory of scientific growth. **282** (1978).

46. J. Ziman, *Real Science: What It Is and What It Means* (Cambridge University Press, Cambridge, England, 2000).

47. M. Strevens, *The Knowledge Machine: How Irrationality Created Modern Science* (Liveright Publishing Corporation, New York, NY, 2020).

48. H. W. de Regt, *A Contextual Theory of Scientific Understanding* (Oxford University Press, 2017).

49. J. Schumpeter, Creative destruction. *Capitalism, socialism and democracy* **825**, 82–85 (1942).

50. F. Xu, L. Wu, J. Evans, Flat Teams Drive Scientific Innovation. *Proc. Natl. Acad. Sci. U. S. A.* **119** (2022).

51. Y. Lin, C. B. Frey, L. Wu, Remote collaboration fuses fewer breakthrough ideas. *Nature* **623**, 987–991 (2023).

52. C. Davies, B. Page, C. Driesener, Z. Anesbury, S. Yang, J. Bruwer, The power of nostalgia: Age and preference for popular music. *Mark. Lett.* **33**, 681–692 (2022).

53. C. Danescu-Niculescu-Mizil, R. West, D. Jurafsky, J. Leskovec, C. Potts, "No country for old members: User lifecycle and linguistic change in online communities" in *Proceedings of the 22nd International Conference on World Wide Web* (2013), pp. 307–318.

54. C. Candia, B. Uzzi, Quantifying the selective forgetting and integration of ideas in science and technology. *Am. Psychol.* **76**, 1067–1087 (2021).

55. L. Li, Y. Lin, L. Wu, Innovation by Displacement, *arXiv [cs.DL]* (2025). https://doi.org/10.48550/arXiv.2506.15959.

56. A. Sinha, Z. Shen, Y. Song, H. Ma, D. Eide, B.-J. (paul) Hsu, K. Wang, "An Overview of Microsoft Academic Service (MAS) and Applications" in *Proceedings of the 24th International Conference on World Wide Web* (Association for Computing Machinery, New York, NY, USA, 2015)*WWW '15 Companion*, pp. 243–246.

57. Y. Xing, Y. Ma, Y. Fan, R. Sinatra, A. Zeng, Academic mentees thrive in big groups, but survive in small groups. *Nat. Hum. Behav.*, doi: 10.1038/s41562-025-02114-8 (2025).

58. M. Andalón, C. de Fontenay, D. K. Ginther, K. Lim, The rise of teamwork and career prospects in academic science. *Nat. Biotechnol.* **42**, 1314–1319 (2024).

59. S. F. Way, A. C. Morgan, D. B. Larremore, A. Clauset, Productivity, prominence, and the effects of academic environment. *Proc. Natl. Acad. Sci. U. S. A.* **116**, 10729–10733 (2019).

60. P. Azoulay, J. S. Graff Zivin, B. N. Sampat, The Diffusion of Scientific Knowledge Across Time and Space: Evidence from Professional Transitions for the Superstars of Medicine (2011). https://doi.org/10.3386/w16683.

61. D. Jurgens, S. Kumar, R. Hoover, D. McFarland, D. Jurafsky, Measuring the evolution of a scientific field through citation frames. *Trans. Assoc. Comput. Linguist.* **6**, 391–406 (2018).

62. C. Catalini, N. Lacetera, A. Oettl, The incidence and role of negative citations in science. [Preprint] (2015). https://doi.org/10.1073/pnas.1502280112.

63. D. J. Price, NETWORKS OF SCIENTIFIC PAPERS. *Science* **149**, 510–515 (1965).

64. K. Kedrick, E. Levitskaya, R. J. Funk, Conceptual structure and the growth of scientific knowledge. *Nat. Hum. Behav.* **8**, 1915–1923 (2024).

65. S. Milojević, Quantifying the cognitive extent of science. *J. Informetr.* **9**, 962–973 (2015).

66. A. M. Petersen, F. Arroyave, F. Pammolli, The Disruption Index Suffers From Citation Inflation and Is Confounded by Shifts in Scholarly Citation Practice (2023). https://doi.org/10.2139/ssrn.4486421.

67. M. Park, E. Leahey, R. J. Funk, Papers and patents are becoming less disruptive over time. *Nature* **613**, 138–144 (2023).

68. E. E. Salas, S. M. Fiore, *Team Cognition: Understanding the Factors That Drive Process and Performance* (American Psychological Association, 2004).

69. A. W. Woolley, C. F. Chabris, A. Pentland, N. Hashmi, T. W. Malone, Evidence for a collective intelligence factor in the performance of human groups. *Science* **330**, 686–688 (2010).

70. J. S. G. Chu, J. A. Evans, Slowed canonical progress in large fields of science. *Proc. Natl. Acad. Sci. U. S. A.* **118** (2021).

71. J. A. Evans, Electronic publication and the narrowing of science and scholarship. *Science* **321**, 395–399 (2008).

72. V. Larivière, É. Archambault, Y. Gingras, "Long-term patterns in the aging of the scientific literature, 1900--2004" in *Proceedings of the 11th Conference of the International Society for Scientometrics and Informetrics (ISSI), Madrid, Spain* (2007), pp. 449–456.

73. D. Adams, How to Stop Worrying and Learn to Love the Internet (1999). https://douglasadams.com/dna/19990901-00-a.html.

74. D. W. Galenson, Painting outside the lines: Patterns of creativity in modern art. (2001).

75. D. W. Galenson, *Old Masters and Young Geniuses: The Two Life Cycles of Artistic Creativity* (Princeton University Press, Princeton, NJ, 2007).

76. M. B. Holbrook, R. M. Schindler, Some exploratory findings on the development of musical tastes. *J. Consum. Res.* **16**, 119 (1989).

77. Q. Ke, E. Ferrara, F. Radicchi, A. Flammini, Defining and identifying Sleeping Beauties in science. *Proc. Natl. Acad. Sci. U. S. A.* **112**, 7426–7431 (2015).

78. A. F. J. van Raan, Sleeping Beauties in science. *Scientometrics* **59**, 467–472 (2004).

79. C. Franzoni, G. Scellato, P. Stephan, Foreign-born scientists: mobility patterns for 16 countries. *Nat. Biotechnol.* **30**, 1250–1253 (2012).

80. NCSES, “The STEM Labor Force of Today: Scientists, Engineers, and Skilled Technical Workers” (NSB-2021–2, 2021); https://ncses.nsf.gov/pubs/nsb20212.

81. C. Franzoni, P. Stephan, R. Veugelers, Funding Risky Research. *Entrepreneurship and Innovation Policy and the Economy* **1**, 103–133 (2022).

82. C. Haeussler, H. Sauermann, Division of labor in collaborative knowledge production: The role of team size and interdisciplinarity. *Res. Policy* **49**, 103987 (2020).

83. R. B. Freeman, W. Huang, Collaborating with people like me: Ethnic coauthorship within the United States. *J. Labor Econ.* **33**, S289–S318 (2015).

84. C. R. Sugimoto, N. Robinson-García, D. S. Murray, A. Yegros-Yegros, R. Costas, V. Larivière, Scientists have most impact when they’re free to move. *Nature* **550**, 29–31 (2017).

48. Price, D. J. Citation measures of hard science, soft science, technology, and nonscience. *Communication among scientists and engineers* 3, 22 (1970).

49. Larivière, V., Archambault, É. & Gingras, Y. Long-term patterns in the aging of the scientific literature, 1900--2004. in *Proceedings of the 11th conference of the international society for scientometrics and informetrics* (ISSI), Madrid, Spain 449–456 (2007).